\documentclass[sigconf]{acmart}
\AtBeginDocument{%
  \providecommand\BibTeX{{%
    \normalfont B\kern-0.5em{\scshape i\kern-0.25em b}\kern-0.8em\TeX}}}

\pdfoutput=1

\setcopyright{acmcopyright}
\copyrightyear{2023}
\acmYear{2023}
\acmDOI{XXXXXXX.XXXXXXX}

\acmConference[ICSE'24]{46th International Conference on Software Engineering (ICSE 2024)}{April 14-20, 2024}{Lisbon, Portugal}
\acmBooktitle{ICSE'24:46th International Conference on Software Engineering (ICSE 2024) April 14-20, 2024, Lisbon, Portugal} 
\acmPrice{}
\acmISBN{}

\usepackage{enumitem}
\usepackage{hyperref}
\usepackage{multirow}
\usepackage{booktabs}
\begin{document}

\title{Exploring the Need of Accessibility Education in the Software Industry: Insights from a Survey of Software Professionals in India}

%% Of note is the shared affiliation of the first two authors and the
 
%%\author{P D Parthasarathy}\email{p20210042@goa.bits-pilani.ac.in}\orcid{0000-0002-8723-2407}
%%\affiliation{%
%%\institution{BITS Pilani, KK Birla Goa Campus}
%%\city{}\state{Goa}
%%\country{India}
%%\postcode{403726}
%%}
%%\author{Swaroop Joshi}\email{swaroopj@goa.bits-pilani.ac.in}\orcid{0000-0003-4536-2446}
%%\affiliation{%
%%  \institution{BITS Pilani, KK Birla Goa Campus}
%%  \city{}\state{Goa}
%%  \country{India}
%%  \postcode{403726}
%%}
%%\author{Dev Goel}\email{f20190236@goa.bits-pilani.ac.in}\orcid{0009-0006-4775-4776}
%%\affiliation{%
%%  \institution{BITS Pilani, KK Birla Goa Campus}
%%  \city{}\state{Goa}
%%  \country{India}
%%  \postcode{403726}
\author{P D Parthasarathy}\email{p20210042@goa.bits-pilani.ac.in}\orcid{0000-0002-8723-2407}
\affiliation{%
\institution{BITS Pilani, KK Birla Goa Campus}
\city{}\state{Goa}
\country{India}
\postcode{403726}
}
\author{Swaroop Joshi}\email{swaroopj@goa.bits-pilani.ac.in}\orcid{0000-0003-4536-2446}
\affiliation{%
 \institution{BITS Pilani, KK Birla Goa Campus}
 \city{}\state{Goa}
 \country{India}
 \postcode{403726}
}
% \author{First author}\email{email@email}\orcid{0000-0000-0000-0000}
% \affiliation{%
%   \institution{Univerisity Name}
%   \city{}\state{State}
%   \country{Country}
%   \postcode{Code}
% }
% \author{Second author}\email{email@email}\orcid{0000-0000-0000-0000}
% \affiliation{%
%   \institution{Univerisity Name}
%   \city{}\state{State}
%   \country{Country}
%   \postcode{Code}
% }

\renewcommand{\shortauthors}{}

\begin{abstract}
    
A UserWay study in 2021 indicates that an annual global e-commerce revenue loss of approximately \$16 billion can be attributed to inaccessible websites and applications. According to the 2023 WebAIM study, only 3.7\% of the world's top one million website homepages are fully accessible. This shows that many software developers use poor coding practices that don't adhere to the Web Content Accessibility Guidelines (WCAG). This research centers on software professionals and their role in addressing accessibility. This work seeks to understand (a) who within the software development community actively practices accessibility, (b) when and how accessibility is considered in the software development lifecycle, (c) the various challenges encountered in building accessible software, and (d) the resources required by software professionals to enhance product accessibility. Our survey of 269 software professionals from India sheds light on the pressing need for accessibility education within the software industry. A substantial majority (69.9\%, N=269) of respondents express the need for training materials, workshops, and bootcamps to enhance their accessibility skills. We present a list of actionable recommendations that can be implemented within the industry to promote accessibility awareness and skills. We also open source our raw data for further research, encouraging continued exploration in this domain.
 
%While similar studies have been conducted by researchers and organizations like WebAIM and LevelAccess, they provide limited representation of the Indian IT sector, which hosts a high incidence of disabilities and a substantial population of software professionals.
\end{abstract}




\newenvironment{myquote}%
  {\list{}{\leftmargin=0.1in\rightmargin=0.1in}\item[]}%
  {\endlist}

\maketitle
% !TeX root = main.tex
\section{Introduction}
Digital accessibility involves designing and developing digital artifacts like websites, applications, documents, and online content such that they can be accessed, understood, and used by a wide range of individuals, including those with disabilities. The objective of digital accessibility is to ensure equal access and opportunities for people with disabilities, allowing them to participate in the digital world alongside individuals without disabilities. Digital accessibility is not optional; rather, it is a crucial, implicit, non-functional requirement of all software. 

Accessibility is mandated by various governments, including the United States \cite{Section508}, European Union \cite{EU_web_2023}, Canada \cite{secretariat_standard_2011}, and India \cite{ministryofsocialjusticeandempowermentgovernmentofindiaRightsPersonsDisabilities2016}. Industry-specific accessibility standards such as the 21st Century Communications and Video Accessibility Act (CVAA) \cite{cvaa_21st_2021} and Air Carrier Access Act \cite{accAct} are on the rise. However, according to the 2023 WebAIM report\footnote{\url{https://webaim.org/projects/million/}}, just 3.7\% of the world's top one million website homepages provide full accessibility. Meanwhile, nearly 97\% of around 500 popular Android apps spanning 23 business categories show accessibility violations \cite{yan_current_2019}. As per the 2021 UserWay report\footnote{\url{https://s3.amazonaws.com/media.mediapost.com/uploads/Userway_Report.pdf}}, inaccessible applications and websites cause a global revenue loss of around \$16 billion in the e-commerce sector alone. Therefore, inaccessibility is not only a matter of social justice but also results in business losses and legal challenges.

Several automated testing tools, such as Axe\footnote{\url{https://www.deque.com/axe/}} and Wave\footnote{\url{https://wave.webaim.org/}}, can assess WCAG compliance. It's important to note that these tools may not cover all \textit{success criteria} or guidelines comprehensively \cite{pool_accessibility_2023}. Additionally, they can only identify potential violations and cannot fix the issues automatically. Software engineers must address these issues by fixing them manually. Research indicates that no single tool provides comprehensive coverage; each tool has its advantages and disadvantages, reporting varying categories of accessibility violations \cite{pool_accessibility_2023}. Furthermore, these tools may generate both false positives and false negatives. On the other hand, the utilization of \textit{accessibility overlays}, despite their longstanding presence, has been a subject of scrutiny and disapproval from accessibility experts\footnote{\url{https://www.levelaccess.com/blog/the-criticisms-and-objections-of-accessibility-overlays/}}. These overlays introduce an extra layer of code to existing software with the aim of making the software WCAG compliant. However, as overlays do not repair underlying issues, they don’t actually help software products achieve complete accessibility. To build accessible software products and services, the W3C recommends integrating accessibility considerations into the software development process as early as possible \cite{w3cStartwitha11y}.

The literature emphasizes the human element, namely, the software engineer and their role in addressing accessibility concerns. This focus is well-founded because it is the software developer who is responsible for coding, introducing potential software inaccessibility into the codebase, and, ultimately, interpreting the results from accessibility automation tools. Furthermore, it will also fall upon the software developer to rectify any accessibility violations in the code. Patel et al. \cite{patel_why_2020} conducted a survey involving 77 technology professionals to explore their perspectives and challenges related to accessibility. They found that formal education did not adequately prepare these professionals to address accessibility topics. WebAIM and LevelAccess, in partnership with the Global Initiative for Inclusive ICTs (G3ICT) and the International Association of Accessibility Professionals (IAAP), have undertaken a series of studies to assess the state of digital accessibility among technology professionals. Notably, these studies primarily featured respondents from the Global North, with less than 5\% respondents from the Global South.

Nevertheless, it's crucial to recognize the substantial impact that the Global South, particularly India, has on software development. As per StackOverflow's 2022 developer survey\footnote{\url{https://survey.stackoverflow.co/2022}}, India is the second-largest developer community globally, following only the United States. Moreover, projections from the 2019 Evans Data Corp's Global Developer Population and Demographic Study indicate that India is expected to surpass the United States in terms of developer population by 2024. According to the 2023 strategic review report by the National Association of Software and Service Companies (NASSCOM) on the technology sector in India \cite{nasscom2023}, approximately 5.4 million software professionals are working in around 35,000 technology firms. India holds a significant 57\% share in the global sourcing of software and is home to about 27,000 startups in the technology sector. This underlines India's crucial role in the Global software industry. 

However, a study by the Economic Times in August 2023\footnote{\url{https://economictimes.indiatimes.com/jobs/hr-policies-trends/top-indian-companies-have-very-few-people-with-disabilities-on-rolls/articleshow/102753098.cms}}, reveals that India's top corporates have only three persons with disabilities (PwDs) for every 1000 employees. Barriers identified include a scarcity of PwD candidates with the required credentials and companies struggling to provide accessibility and a welcoming work environment for PwDs. Additionally, there are notable dynamics concerning accessibility, disabilities, and inclusion in India. For example, the enrollment of students with disabilities in higher education is extremely low (0.19\%) \cite{department_of_higher_education_government_of_india_all_2021}, and unlike the Global North, very few institutions have disability services offices on their campuses. This scarcity suggests that awareness about accessibility among computer science teachers, students, and software professionals in this region, is likely to be limited. 

Our main goal in undertaking this study is to explore accessibility awareness and technical skills among software professionals by addressing the question, \textit{Is there a need for accessibility education in the Indian software industry?}. We conducted an extensive survey in India with the goal of addressing the following research questions:
\begin{itemize}
    \item \textbf{RQ1:} Who within the technology profession practices accessibility? 
    \item \textbf{RQ2:} How and when do software professionals consider and implement accessibility?
    \item  \textbf{RQ3:} What are the challenges encountered in ensuring accessibility? 
    \item  \textbf{RQ4:} What resources are required by software professionals to ensure accessibility? 
\end{itemize}
This paper is organized as follows: Sec \ref{sec:relatedwork} describes the related literature, and Sec \ref{sec:method} explains the survey methodology. The findings are presented in Sec \ref{sec:findings} and discussed subsequently in Sec \ref{sec:discussion}. Sec \ref{sec:Recommendations } lists actionable recommendations for accessibility education in the technology industry. Sec \ref{sec:threats} discusses the threats to validity and the impact of this work. Sec \ref{sec:conclusion} concludes our study.  
% !TeX root = main.tex
\section{Related Work}
\label{sec:relatedwork}
Previous research has explored the teaching of accessibility in higher education; A 2020 literature review by Baker et al. on teaching accessibility in CS curricula \cite{baker_systematic_2020} identifies four broad learning objectives of teaching accessibility: (a) Accessibility Awareness - abilities, laws, and ethics, (b) Technical Knowledge - guidelines, ways to implement accessibility features, testing, (c) Empathy, and (d) Career options in accessibility. Several authors have reported on teaching accessibility in college-level computing courses such as web-Designing \cite{kawasTeachingAccessibilityDesign2019}, Software Engineering \cite{el-glalyTeachingAccessibilitySoftware2020}, CS1/CS2 \cite{jiaInfusingAccessibilityProgramming2021}, Mobile Application development \cite{bhatia_integrating_2023}, and AI \cite{tseng_exploration_2022} using conventional classroom modules or using project-based learning.

While academia is taking these steps to equip computing students with accessibility skills, there is limited knowledge regarding how technology professionals incorporate accessibility into their work and utilize resources to acquire accessibility knowledge while on the job. Research investigating web accessibility since 1999 has observed positive developments. Yet, these observations suggest that improvements in web browsers, a focus on search engine page rankings, and shifts in coding standards might have exerted a more significant influence on the outcomes than a specific emphasis on accessibility \cite{richards_web_2012}. Patel et al. \cite{patel_why_2020} conducted a survey involving 77 technology professionals from the United States to explore their perspectives and challenges related to accessibility. The findings from their study suggest that professionals lacked formal accessibility expertise, faced limitations in tools and resources, and overlooked retroactive adjustments in project timelines. While their emphasis is on exploring challenges in integrating accessibility into technology solutions and understanding the tools and resources utilized or needed by technology professionals for creating accessible solutions, our focus extends beyond these aspects. We aim to comprehend how accessibility is integrated into the SDLC and identify the stakeholders involved in ensuring accessibility, along with understanding the barriers to ensuring accessibility and the various resources needed to ensure accessibility.

While the 2023 NASSCOM report \cite{nasscom2023} and 2022 GitHub reports \footnote{\url{https://octoverse.github.com/2022/global-tech-talent}} indicate that Asia is home to over 6.5 million software engineers, with over 5 million in India, a survey conducted by WebAIM in January 2021 \cite{webAim_Pract_Survey} has very minimal respondents from Asia. In this survey of web accessibility practitioners, only 2.8\% of respondents (N=757) were from Asia. The findings indicated that over half of the participants believed that accessibility had either not improved or remained unchanged over the past few years. Furthermore, the survey revealed that 91.3\% of respondents acquired their accessibility knowledge through online resources, while 81.1\% learned by collaborating with peers and colleagues. It's worth noting that this study specifically targeted accessibility practitioners, although there is growing advocacy for the idea that accessibility is a collective responsibility and that a \textit{shift-left} approach should be adopted to integrate accessibility into the software development life cycle (SDLC) \cite{armas_proposal_2020}. In the shift-left approach to accessibility, accessibility is considered early in the design and development phases rather than being deferred until the later stages of development or testing stages of the product. It also emphasizes the philosophy that everyone, from designers, developers, and management, plays a role in accessibility, and it is not solely the responsibility of testers to ensure accessibility. 

LevelAccess, in collaboration with the Global Initiative for Inclusive ICTs (G3ICT) and the International Association of Accessibility Professionals (IAAP), conducted an extensive survey in 2022 to gain insights into the status of digital accessibility \cite{LevelAccess2022}. Among the respondents, 76.8\% (N=1,030) were from the United States, while 23.2\% (N=1,030) represented the rest of the world. In terms of challenges, 51.5\% of those surveyed expressed difficulties acquiring training on accessibility best practices within their accessibility programs. Interestingly, a significant portion of the developers, specifically 71.6\%, reported that accessibility was a stipulated requirement in their most recent project. However, a noteworthy 36.6\% of these individuals admitted to not fully grasping how to fulfill these accessibility requirements. Additionally, 46.4\% of respondents rated their development team's familiarity with accessibility as "Elementary." Martin et al. \cite{martin_landscape_2022} delved into the demand for accessibility skills among software designers and developers by examining job postings on LinkedIn. Their findings revealed that a significant portion of these job listings did not include any requirements related to accessibility skills. Interestingly, they also noted that for many positions specifically focused on accessibility, educating developers and designers about accessibility was a mandatory requirement. 

Our study adds to the existing body of literature by presenting the first set of results that exclusively address the research questions within the context of India, which is a house for millions of software professionals. In addition to exploring the barriers and resources necessary for ensuring accessibility, we investigate who actively participates in accessibility practices and the methods and stages at which accessibility is implemented.

% !TeX root = main.tex
\section{Method}
\label{sec:method}
An online questionnaire was utilized to conduct a survey with the aim of gaining deeper insights for our research questions RQ1 through RQ4. The digital format of the survey facilitated the ability to reach and survey a substantial population of software professionals.
\subsection{Survey Design}
The survey commenced by presenting a concise explanation of the concept of digital accessibility, accompanied by the informed consent form approved by the institute’s Human Ethical Committee (HEC). Only those who signed it could proceed to the actual instrument. The core part of the survey is reproduced below for the reader's benefit:

\begin{enumerate}
    \item How would you rate your knowledge in the field of digital accessibility? \textsc{ [A 5-point rating Likert scale, 5 being highest]} \label{label:selfRating}
    \item Mention your work domain. \textit{\textsc{[Select One:}} \textit{(a) User Research}, \textit{(b) User Experience (UX) Design}, \textit{(c) Software Development}, \textit{(d) Software Testing}, \textit{(e) Software Operations (DevOps, Scrum Master etc.)}, \textit{(f) Software Security}, \textit{(g) Software Management (Lead, Manager, Director etc.)}, \textit{(h) Human Resources (HR)}, \textit{(i) Customer support}, \textit{(j) Other (Specify)]}
    \item How many years of experience do you have in the software industry? \textit{ \textsc{[Select one:} (a) 0-3 years, (b) 4-7 years, (c) 8-15 years, (d) 15+ years]}
    \item Do you work/have worked in the past in an accessibility-related position/role in your company? If yes, please mention your role. Otherwise, mention "No". \textsc{[Short text answer]} \label{label:Q4}
    \item Is your work location based out of India? \textsc{[Select One]: [Yes, No]} 
    \item Which of the following categories best describes your company? \textit{\textsc{[Select One:} (a) Product based, (b) Service based, (c) Hybrid (companies which work in both product and service), (d) Other (Specify)]} \label{label:Q6}
    \item Which of the following categories best describes the size of your company? \textit{\textsc{[Select One:} (a) Small scale (1-100 employees), (b) Medium scale (101-1000 employees), (c) Large scale (1000+ employees)]} 
    \item What is your company's business model? \textit{\textsc{[Select One:} (a) B2B / Business-to-Business model, (b) B2C / Business-to-Consumer model, (c) C2C / Consumer-to-Consumer model, (d) C2B / Consumer-to-Business model]} 
    \item Does your company work towards developing assistive technology? (e.g., developing screen readers or gadgets like braille keyboards, switches, etc.) \textit{\textsc{[Select One:} (a) Yes, (b) No, (c) Maybe]} \label{label:Q9}
    \item Your company provides sufficient resources for making its products accessible? \textsc{[A 5-point Agree-Disagree Likert scale]} \label{label:Q10}
    \item Have you worked on any project/assignment that involved inclusive design practice in your company? \textit{(Inclusive design is a design process in which a product, service, or environment is designed to be usable for as many people as possible, particularly groups who are traditionally excluded from being able to use an interface or navigate an environment. Inclusive design describes methodologies to create products that understand and enable people of all backgrounds and abilities. It may address accessibility, age, economic situation, geographic location, language, race, and more. Inclusive design is a process, and accessibility is an outcome.)} \textsc{[Select One]: [Yes, No]} 
    \begin{itemize}
        \item If yes, go to the next question, Otherwise, go to Q. ~\ref{qn:all}
    \end{itemize}
    \item Was accessibility testing performed in any project you worked on in your company? Choose the most appropriate option from the list below. \textit{\textsc{[Select One:} (a) Accessibility testing was done but was outsourced, (b) We had a dedicated team for handling accessibility, (c) Did not perform accessibility testing]}
    \item Was user testing performed in any project you worked on in your company? Choose the most appropriate option from the list below. \textit{\textsc{[Select One:} (a) User testing was done, and we carefully picked users to include people with disabilities as well, (b) User testing was done, but users were chosen at random (c) Did not perform user testing]}
    \item At what stages was inclusive design practiced (and accessibility considerations made)? \textit{\textsc{[Multi-select:} (a) Designing stage of the product, (b) Development stage, (c) Testing stage, (d) I am not sure]} \label{label:Q14}
    \item Which accessibility standards were followed while developing the software? \textit{\textsc{[Multi-select:} (a) Web Content Accessibility Guidelines (WCAG 2.1 or above), (b) User Agent Accessibility Guidelines (UAAG), (c) Authoring Tool Accessibility Guidelines (ATAG), (d) No accessibility standards were followed, (e) Other (specify)]} 
    \item Did the development team include anyone who had relevant accessibility expertise (experience with assisstive technology, Section 508 Trusted Tester certified, IAAP certified etc.) to provide inputs to accessibility-related aspects of the project? \textit{\textsc{[Select One:} (a) Yes, my team had someone who had enough expertise to provide relevant input to the accessibility design of the project/accessibility-related aspects of the project, (b) No, my team had nobody who could provide relevant input, (c) I was not a part of the team}
    \item  $\triangleright$ What are the challenges industry professionals face in incorporating accessibility into their software product/service? \label{qn:all} \textit{\textsc{[Multi-select:} (a) Lack of skilled developers/professionals, (b) No recognized parameters to certify a product/service as accessible, (c) Difficult to recruit people with disabilities with technical knowledge, (d) Difficult to accommodate people with disabilities, (e) Lack of awareness, (f) Other (Specify)]} \label{label:Q17}
    \item Please specify what other resources you would want to facilitate the process of making accessible software products. \textsc{[Short-Answer]} \label{label:Q18}
\end{enumerate}
In addition to the above-mentioned questions, the survey encompassed general inquiries about participants' gender, qualifications, whether they identify as a PwD, and whether they have any PwD in their immediate family or professional network. The survey underwent a pilot run with industry colleagues and alumni from our research lab currently employed in the software industry to evaluate the clarity of questions and the overall duration of the survey.

\subsection{Recruiting Participants }
A purposive sample was employed to recruit participants, chosen for its simplicity and effectiveness. The survey utilized Google Forms and was disseminated through email to the authors' professional connections within the Indian software industry. The members of the NASSCOM\footnote{\url{https://nasscom.in/}} were considered to ensure a diverse representation; professionals from Product-based, Service-based, Hybrid, start-ups, as well as medium and large-sized organizations were targeted. Additionally, the survey link was shared with alumni who had graduated with a computing major at our institution. In total, 14,339 emails were sent, and the survey remained open for participation from February 10, 2023, to March 31, 2023.
% !TeX root = main.tex
\section{Findings}
\label{sec:findings}
A total of 269 responses were received. The responses collected from the Google form were safely stored in cloud storage for further analysis. We applied quantitative analysis to answer RQ1 - RQ3 and qualitative analysis; in particular, structural coding to categorize and count their frequencies as in described by Saldana (The Coding Manual for Qualitative Researchers, Ch 3) \cite{saldana_coding_2021} to answer RQ4. We analyzed the participant profile, and the results are presented in Table \ref{table:participantSummary}.  

\begin{table}[h]
\vspace{-0.4cm}
\caption{\centering Participant Profile Summary (N=269)}\label{table:participantSummary}
\vspace{-10pt}
\centering
        \begin{tabular}{lll}
            \toprule
            \multirow{1}{*}{\textbf{Cateogry}} & \multicolumn{1}{l}{\textbf{Sub-Category}} & \multicolumn{1}{l}{\textbf{Percentage}}
            \\\hline
            \multirow{3}{*}{Gender} & \multicolumn{1}{l}{Male} & \multicolumn{1}{l}{59.5\%} \\\cline{2-3}
                                 & \multicolumn{1}{l}{Female} & \multicolumn{1}{l}{23.4\%} \\\cline{2-3}
                                 & \multicolumn{1}{l}{Not Disclosed} & \multicolumn{1}{l}{17.1\%} \\\hline
            \multirow{3}{*}{Qualification} & \multicolumn{1}{l}{Bachelor's} & \multicolumn{1}{l}{62.8\%} \\\cline{2-3}
                                 & \multicolumn{1}{l}{Master's} & \multicolumn{1}{l}{36.4\%}  \\\cline{2-3}
                                 & \multicolumn{1}{l}{Ph.D.} & \multicolumn{1}{l}{0.7\%} \\\hline 
            \multirow{4}{*}{Experience} & \multicolumn{1}{l}{0-3 Years} & \multicolumn{1}{l}{31.2\%} \\\cline{2-3}
                                 & \multicolumn{1}{l}{4-7 Years} & \multicolumn{1}{l}{27.5\%} \\\cline{2-3}
                                 & \multicolumn{1}{l}{8-15 Years} & \multicolumn{1}{l}{23.8\%} \\\cline{2-3}
                                 & \multicolumn{1}{l}{15+ Years} & \multicolumn{1}{l}{17.5\%} \\\hline
            \multirow{6}{*}{Domain } & \multicolumn{1}{l}{Software Developer/Tester} & \multicolumn{1}{l}{56.6\%} \\\cline{2-3}
                                 & \multicolumn{1}{l}{Software Management} & \multicolumn{1}{l}{16.0\%} \\\cline{2-3}
                                 & \multicolumn{1}{l}{Software Operations} & \multicolumn{1}{l}{7.8\%} \\\cline{2-3}
                                  & \multicolumn{1}{l}{Software UX Design} & \multicolumn{1}{l}{6.7\%} \\\cline{2-3}
                                   & \multicolumn{1}{l}{Customer Support} & \multicolumn{1}{l}{2.2\%} \\\cline{2-3}
                                 & \multicolumn{1}{l}{Others} & \multicolumn{1}{l}{10.7\%} \\\hline                  \multirow{3}{*}{Identify as PwD} & \multicolumn{1}{l}{Yes} & \multicolumn{1}{l}{4.08\%} \\\cline{2-3}
                                 & \multicolumn{1}{l}{No} & \multicolumn{1}{l}{80.29\%} \\\cline{2-3}
                                 & \multicolumn{1}{l}{Prefer not to say} & \multicolumn{1}{l}{15.61\%}  \\\hline
        \end{tabular}\vspace{-8pt}
\end{table}
The participants of the survey represented a diverse group of organization types. Table \ref{table:participantOrgSummary} describes the respondent's organization-related findings.

\begin{table}[h]
\caption{\centering Participants' Organization Summary (N=269)}\label{table:participantOrgSummary}
\vspace{-10pt}
\centering
        \begin{tabular}{lll}
            \toprule
            \multirow{1}{*}{\textbf{Cateogry}} & \multicolumn{1}{l}{\textbf{Sub-Category}} & \multicolumn{1}{l}{\textbf{Percentage}}
            \\\hline
             \multirow{3}{*}{Company Size} & \multicolumn{1}{l}{Large scale} & \multicolumn{1}{l}{71.7\%} \\\cline{2-3}
                                 & \multicolumn{1}{l}{Medium scale } & \multicolumn{1}{l}{23.4\%} \\\cline{2-3}
                                 & \multicolumn{1}{l}{Small scale } & \multicolumn{1}{l}{4.8\%}  \\\hline
            \multirow{3}{*}{Company Modal} & \multicolumn{1}{l}{B2B} & \multicolumn{1}{l}{50.6\%} \\\cline{2-3}
                                 & \multicolumn{1}{l}{B2C} & \multicolumn{1}{l}{37.5\%} \\\cline{2-3}
                                 & \multicolumn{1}{l}{C2C} & \multicolumn{1}{l}{3.0\%} \\\cline{2-3} 
                                & \multicolumn{1}{l}{Others} & \multicolumn{1}{l}{8.9\%} \\\hline
            \multirow{3}{*}{Company Type} & \multicolumn{1}{l}{Service Based } & \multicolumn{1}{l}{34.2\%} \\\cline{2-3}
                                 & \multicolumn{1}{l}{Product based } & \multicolumn{1}{l}{32.7\%} \\\cline{2-3}
                                 & \multicolumn{1}{l}{Hybrid} & \multicolumn{1}{l}{32.0\%} \\\cline{2-3} 
                                & \multicolumn{1}{l}{Others} & \multicolumn{1}{l}{1.1\%} \\\hline
              \multirow{3}{*}{Firm in Assistive Tech} & \multicolumn{1}{l}{Yes} & \multicolumn{1}{l}{19.3\%} \\\cline{2-3}
                                 & \multicolumn{1}{l}{No} & \multicolumn{1}{l}{58.0\%} \\\cline{2-3}
                                 & \multicolumn{1}{l}{Not Sure} & \multicolumn{1}{l}{22.7\%}  \\\hline 
        \end{tabular}
\end{table}

Regarding their proficiency in accessibility skills, 39.8\% of the respondents (N=269) assessed themselves with a score of 1 or 2 on a scale ranging from 1(lowest) to 5(highest). 16.4\% of the respondents provided a neutral rating of 3, and 43.8\% of them rated their accessibility skills as 4 or 5. 

\begin{figure}[H]
\centering
    \includegraphics[scale=0.6]{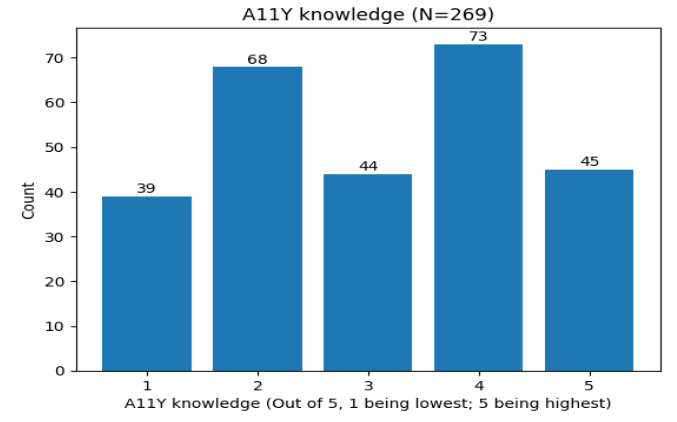}
    \caption{\centering Accessibility Knowledge self-reported by respondents}
    \label{fig:knowledgeSelfReported}
    \vspace{-4mm} %5mm vertical space
\end{figure}

The respondents' average ratings amounted to 3.06, with a standard deviation of 1.33. Figure \ref{fig:knowledgeSelfReported} illustrates the results of Q\ref{label:selfRating}. In the rest of this section, we discuss the findings related to each of our RQs.

\vspace{-8pt}

\subsection{RQ1: Who practices Accessibility ?}
A mere 4.4\% of the respondents (N=269), which translates to only 12 participants, reported having experience in roles related to accessibility in their current or past positions. Out of these 12 individuals, 11 were male, and 1 female and none identified as persons with disabilities. 11 of them belong to a large-scale company and 1 of them belonged to a medium-scale company. Their experience levels varied, with five having over 15 years of experience, three having 8-15 years, three having 0-3 years, and one having 4-7 years of experience. 

Among these individuals, 11 claimed to possess proficient knowledge in accessibility, rating themselves a 4 or 5 on a 5-point scale for accessibility knowledge, while 1 rated themselves as a 3. Regarding job roles, 7 were software engineers, 4 held positions as software leads or managers and 1 was in a customer service and support role. Education-wise, 7 had Master's degrees, while 5 held Bachelor's degrees. Notably, the majority of them worked in B2B (5 individuals) and B2C companies (4 individuals). Six of them were employed in service-based organizations, while five were in product-based organizations (Note that these findings originate from the responses to Q\ref{label:Q6}  to Q\ref{label:Q9}, and the organization’s name was not asked to maintain anonymity). Nine participants indicated that they applied inclusive design principles to ensure accessibility in their work. Two of them reported that their organization was involved in building assistive technology as well. 

\subsection{RQ2:  Accessibility Consideration and Implementation}
Only 39\% of the respondents (N=269), which corresponds to 105 participants, reported that they practiced inclusive design within their organizations to achieve accessibility goals. Among those who implemented inclusive design:
\begin{itemize}
    \item 31.4\% (N=105) mentioned that they did not conduct any accessibility testing.
    \item 21.9\% (N=105) revealed that they outsourced accessibility testing.
    \item 36.2\% (N=105) stated that they performed user testing with individuals having disabilities.
\end{itemize}

\begin{figure}[H]
\centering
    \includegraphics[scale=0.6]{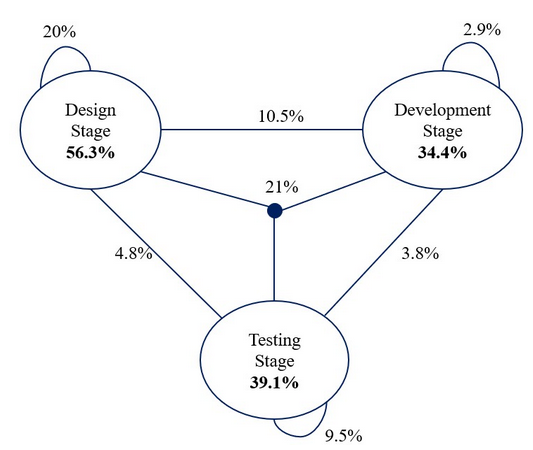}
    \caption{\centering Consideration of a11y in different phases of SDLC}
    \label{fig:a11yconsiderations} \vspace{-8pt}
\end{figure}

Interestingly, 27.5\% (N=105) of the respondents expressed uncertainty regarding when accessibility considerations were incorporated into the software development lifecycle (SDLC). As discussed in Section \ref{sec:relatedwork}, there is an increased emphasis on considering accessibility from \textit{design to deployment} rather than just in the testing phase. In order to understand the a11y considerations across the three major phases of software development (design, development, and testing), we analyzed the responses to Q\ref{label:Q14} whose results are depicted in Figure \ref{fig:a11yconsiderations}. Nodes correspond to the three phases, showing the percentage of respondents (N=105) considering accessibility in that phase. Edges indicate the consideration of accessibility in both its adjacent phases, while self-loops signify exclusive consideration of accessibility in a particular phase.

\vspace{-4mm}

In summary, 
\begin{itemize}
    \item 56.3\% of them incorporated accessibility at the design stage. 
    \item 34.4\% considered accessibility during the development stage.
    \item 39.1\% addressed accessibility during the testing stage.
\end{itemize}

Additionally, 20\% exclusively focused on accessibility during the design stage (and not in other stages), 2.9\% solely during the development stage, and 9.5\% solely during the testing phase. About 21.0\% considered accessibility throughout the entire lifecycle, from design and development to testing. Meanwhile, 10.5\% focused on accessibility during both the design and development phases, 4.8\% during the design and testing phases, and 3.8\% during the development and testing phases.

\begin{figure}[H]
\centering
    \includegraphics[scale=0.6]{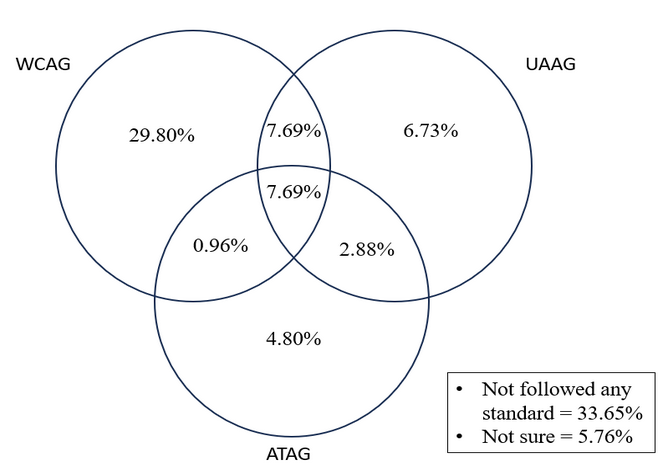}
    \caption{\centering Usage of Accessibility standards }
    \label{fig:standardsFollowed}
\end{figure}

A substantial 33.65\% (N=105) indicated that they did not utilize any accessibility standards, while 5.76\% (N=105) were uncertain about the specific standards in use. On the other hand, 7.69\% reported the utilization of all three standards, namely WCAG, UAAG, and ATAG, while 29.8\% relied solely on WCAG. More detailed information on the usage of these three standards is depicted in Figure \ref{fig:standardsFollowed} as a Venn diagram. 

Among the respondents, only 40.0\% (N=105) reported that their team included individuals with sufficient expertise to contribute to accessibility-related discussions and tasks.

In summary, only a mere 1.9\% (N=105) of the respondents took comprehensive measures such as 
\begin{itemize}
    \item Having a team member with accessibility expertise,
    \item Incorporating accessibility considerations across the SDLC,
    \item Considering all three major accessibility standards such as WCAG, UAAG and ATAG, 
    \item Conducting accessibility testing, and 
    \item Performing user testing involving individuals with disabilities.
\end{itemize}

\subsection{RQ3: Challenges in ensuring accessibility}
Much like the barriers identified in previous research by Patel et al. \cite{patel_why_2020} and Durdu et al. \cite{durdu_perception_2020}, our study also revealed that the primary challenges, consistent with their findings, are the lack of accessibility skills and a general lack of awareness regarding accessibility among software professionals. Table \ref{tab:challenges} provides a list of these challenges, along with their respective percentages. Note that this question allowed respondents to select multiple challenges, so the percentages may not add up to 100.

\begin{table}[thbp]
  \centering
  \caption{Barriers to teaching accessibility\label{tab:challenges}}
  \vspace{-12pt}
  \begin{tabular}{p{6cm}r}
    \toprule
    Challenge    & Percentage \\\midrule
    Lack of skilled developers/professionals & 76.20\% \\
    Lack of awareness & 71.37\% \\
    Difficult to recruit PwDs with technical knowledge & 29.36\%  \\
    Difficult to accommodate PwDs & 15.98\% \\ 
    No recognized parameters to certify a product/service as accessible & 14.49\%  \\ 
    Budgetary concerns to take care of accessibility &  2.60\% \\
    No major return on investment & 0.37\%  \\\bottomrule
  \end{tabular}\vspace{-10pt}
\end{table}
Note that the last two rows in Table \ref{tab:challenges} are challenges mentioned by respondents' apart from the provided options in Q\ref{label:Q17} using the \textit{Other} option. In addition to these well-established top two challenges in previous studies \cite{patel_why_2020,durdu_perception_2020,webAim_Pract_Survey}, our study uncovered specific issues in the Indian context that have not received significant attention in the existing literature. These include the difficulties associated with recruiting persons with disabilities (PwD) with technical knowledge and the challenges in providing necessary accommodations for PwDs.

\subsection{RQ4: Resources required for accessibility}
\begin{figure}[H]
\centering
    \includegraphics[scale=0.30]{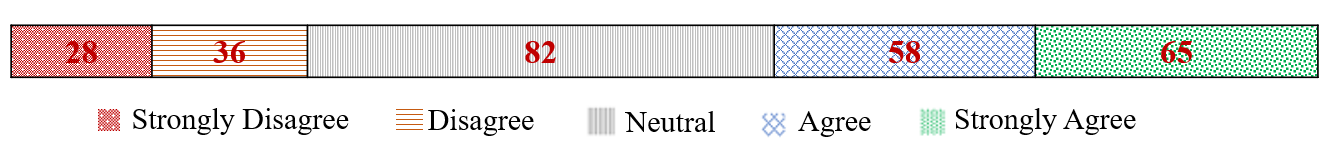}
    \caption{\centering Sufficent resources provided by Organization}
    \label{fig:resourcesAvail}
\end{figure}

A total of 54.2\% of the respondents (N=269) strongly disagreed or disagreed or remained neutral when asked about their organization's support in facilitating resources for ensuring accessibility (Q\ref{label:Q10}). Figure \ref{fig:resourcesAvail} provides an overview of the likert scale responses regarding the provision of accessibility resources by the respondents' organizations.

The open-ended question Q\ref{label:Q18} in our survey aimed to gather insights into the resources professionals considered essential for enhancing the accessibility of software products. To analyze the responses, structural coding was employed, assigning categories to text segments. Through this process, six distinct categories emerged. Table \ref{tab:resources} presents these categories along with the percentage of mentions.

\begin{table}[thbp]
  \centering
  \caption{Resources required for improving accessibility\label{tab:resources}}
  \vspace{-10pt}
  \begin{tabular}{p{6cm}r}
    \toprule
    Resource Required   & Percentage \\\midrule
    Education \& Training & 69.90\% \\
    Budget and Enough Time for Accessibility & 11.65\% \\
    Improved tools and Processes & 6.79\%  \\
    Recruit skilled engineers with a11y knowledge & 3.88\% \\ 
    Facilitate interaction with PwD for user testing  & 2.91\%  \\ 
    Others (Analytics on a11y, better user research) &  5.82\%   \\\bottomrule
  \end{tabular}\vspace{-8pt}
\end{table}

Note that multiple respondents have requested the same resource, so the percentages may not add up to 100. To explore the precise educational resources required by professionals, we employed descriptive coding within the qualitative research framework \cite{saldana_coding_2021}. This method entailed categorizing responses related to education and training into subcategories based on their themes. The outcome yielded six subcategories, as outlined in Table \ref{tab:Edresources}.

\begin{table}[thbp]
  \centering
  \caption{Educational resources required by professionals\label{tab:Edresources}}
  \vspace{-10pt}
  \begin{tabular}{p{6cm}r}
    \toprule
    Educational Resource Required   & Percentage \\\midrule
    Self-learn materials such as books and videos on Accessibility & 36.98\% \\
    Hands-on Tutorials, Workshops and Bootcamps on Accessibility techniques and tools  & 32.87\% \\
    Monthly focused meetings, interaction with a11y experts, regular blogs and articles on happenings in A11Y & 12.32\%  \\
    Sensitize on \textit{why} a11y is important and provide awareness on the field & 10.95\% \\ 
    Facilitate Accessibility Certifications such as ones offered by IAAP   & 5.47\%  \\ 
    Teach Accessibility as part of school or University education &  4.10\%   \\\bottomrule
  \end{tabular}\vspace{-8pt}
\end{table}

% !TeX root = main.tex
\section{Discussion}
\label{sec:discussion}
Although 43.8\% of the participants (N=269) self-reported proficiency in accessibility, there may be a social desirability bias at play. This is suggested by their responses to Q\ref{label:Q4}, where they indicated no prior experience in accessibility-related roles in current or past projects. Notably, the representation of professionals actively engaged in accessibility work is limited (4.4\%, N=269) within the Indian context. 

In line with Patel et al.'s findings \cite{patel_why_2020}, our study also uncovers a similar trend, indicating that small-companies (with size <100) and medium-sized companies (with size <1000) exhibit less activity in accessibility topics, and B2B and B2C organizations demonstrate a stronger emphasis on accessibility compared to other models.

Furthermore, while our study included professionals of various experience levels, a significant majority (66.66\%, N=12) of those reporting involvement in accessibility duties were mid/senior level professionals with 8+ years of experience. This suggests that junior-level professionals may not be sufficiently engaged in accessibility matters. Those who reported undertaking accessibility responsibilities appear to approach it correctly, having engaged in inclusive design and possessing expertise in accessibility.

Additionally, 39\% of respondents practiced inclusive design but fell short of fully implementing its aspects. For example, a significant portion (31.4\%, N=105) of them reported not conducting any accessibility testing, 33.65\% (N=105) did not follow any accessibility standards, and 35.2\% (N=105) indicated a lack of accessibility expertise within their teams. These findings raise concerns about their understanding of accessibility implementation.

The findings reveal that 76.2\% (N=269) and 71.37\% (N=269) of respondents perceive the shortage of engineers with accessibility skills and a lack of awareness as significant barriers to ensuring accessibility which is in line with the results in the US survey conducted by Patel et al. \cite{patel_why_2020}. While academia is taking steps to educate computing students on accessibility, the impact of these efforts may take time to materialize, and there is limited literature on educating the current IT workforce on accessibility. 
Recruiting technically knowledgeable individuals with disabilities (PwDs) to provide insights on accessibility topics across the software development lifecycle (SDLC) is reported as yet another significant challenge. This highlights the broader issue of limited access to education for PwDs, with only a minimal enrollment rate (0.19\%) in higher education in India \cite{department_of_higher_education_government_of_india_all_2021}.

A substantial majority (69.9\%, N=269) of respondents express the need for educational resources, training materials, workshops, and bootcamps to enhance accessibility education. The findings strongly underscore the urgent need for accessibility education within the industry. In the following section, we present recommendations for industry-focused accessibility education, drawing from the findings mentioned in Table \ref{tab:Edresources}, as well as our experience in academia in teaching accessibility and drawing from the first authors' accessibility experience in the software industry.

% !TeX root = main.tex
\section{Recommendations for practitioners}
\label{sec:Recommendations }

\textit{\textbf{\textsc{(1) Customized Training Materials:}}} A notable number of participants, as highlighted in Table \ref{tab:Edresources}, expressed the need for self-learning resources like books, videos, and concise explanations of accessibility concepts. While open-source materials on accessibility topics are available online, this indicates a demand for curated content tailored to their organization. 
    
We suggest incorporating content aligned with the organization's objectives, addressing specific accessibility standards relevant to their sector (e.g., WCAG for web development), or accessibility areas where the company needs improvement. Organization-specific content, such as using examples from their organization's projects and their design systems, ensures that professionals acquire the knowledge and skills pertinent to their roles. Literature also suggests that the learning outcomes in accessibility for a designer are different from those of a software engineer \cite{el-glalyTeachingAccessibilitySoftware2020}. We recommend practitioners curate role-specific materials rather than generic content to teach accessibility.

\textit{\textbf{\textsc{(2) Hands-On Workshops/Bootcamps:}}} Professionals highlight a strong preference for hands-on workshops and bootcamps as the second most sought-after educational resource, as outlined in Table \ref{tab:Edresources}. Our recommendation is to integrate interactive sessions and practical exercises, empowering participants to apply accessibility principles and methods effectively. Utilizing real-world examples and case studies from their own organization can significantly enrich the learning experience. 
    
Additionally, providing training on widely used accessibility testing tools, software, and resources helps participants familiarize themselves and acquire the necessary skills to identify and address accessibility issues efficiently. This hands-on approach ensures professionals are well-equipped for their daily tasks. To further foster accessibility initiatives, identify and nurture champions within organizations based on motivation, participation, and performance in workshops. These champions can serve as advocates and mentors, driving the implementation of accessibility best practices and contributing to establishing a pervasive culture of accessibility within the industry.

\textit{\textbf{\textsc{(3) Events and Newsletters:}}} As emphasized in Table \ref{tab:challenges}, a major barrier to ensuring accessibility is the lack of awareness on accessibility. This is also underscored by the requests for increased interaction with accessibility experts and individuals with disabilities, as detailed in Table \ref{tab:Edresources}. To address this, it is recommended to implement monthly newsletters distributed to all employees within the organization. These newsletters can aim to raise awareness about accessibility topics by featuring the latest updates in the field, compelling business case studies, and exploring the \textit{why} behind accessibility and its significance. 
    
The newsletters should center on practical, real-world scenarios and challenges that professionals may encounter in their daily work. Furthermore, incorporating the perspectives of Persons with Disabilities (PWDs) within the organization is recommended to provide professionals with insights into their lived experiences. If direct involvement of PWDs is not feasible, organizing events such as Global Accessibility Awareness Day (GAAD) and International Day of Disabled Persons can serve as valuable opportunities. These events can include guest speakers, panel discussions, or user testing sessions featuring individuals with disabilities, offering professionals a firsthand understanding of the impact of accessibility through shared lived experiences.

\textit{\textbf{\textsc{(4) Certification and Recognition:}}} As underscored by certain participants, providing industry-recognized certifications or badges upon successfully completing accessibility training programs can serve as an external motivator. We recommend that organizations support their employees in pursuing certifications from reputable bodies such as the International Association of Accessibility Professionals (IAAP) or the Trusted Tester Section 508 by the Department of Homeland Security. 

These certifications not only contribute to career advancement but also signify a dedicated commitment to accessibility. The tangible incentive of certification encourages professionals to engage in accessibility education actively.

\textit{\textbf{\textsc{(5) Budget, Time, and Management Support:}}} Identified as a key resource for ensuring accessibility, participants emphasized the need for sufficient time and budget, as outlined in Table \ref{tab:Edresources}. It is advised that organizational leadership actively supports and prioritizes accessibility education. Encouraging managers and decision-makers to endorse and allocate resources for training programs is crucial. Leadership buy-in plays a vital role in ensuring the success and sustainability of accessibility initiatives.
% !TeX root = main.tex
\section{Threats to Validity \& Contribution}
\label{sec:threats}
\subsection{Threats to Validity}
In this study, we analyze a comprehensive anonymous survey on industry accessibility practices, challenges, and resources required for ensuring accessibility involving 269 participants from India. Due to the online nature of the survey and the anonymization of collected data, control over respondents' actual backgrounds is unattainable. To mitigate potential bias, we ensured that the survey was exclusively sent to professional software developers and clearly stated that it is for software professionals at the survey's commencement. Nevertheless, the uneven representation of organizational sectors (with small-sized companies comprising only 4.8\% of our survey) could introduce bias to our findings and cannot be generalized. 

Our conclusions are drawn from data interpretation, informed by the first author's industry experience. To strengthen the validity, these findings were corroborated by three additional accessibility experts. A Hawthorne effect, where participants may modify behavior due to awareness of being observed, is also possible. As an example, 39.7\% (N=269) of the respondents considered themselves proficient and awarded themselves a rating of 4 or 5 on accessibility for Q\ref{label:selfRating} but have never worked in an accessibility-related role. Similarly, 41.6\% of them (N=269) agreed or strongly that their organization provides them with sufficient resources, whereas they have never worked in an accessibility-related position. For future endeavors, our plan includes conducting interviews with participants who have shown interest in the study to mitigate potential biases. Replicating this work in various regions within the Global South is crucial to validate the generalizability of our findings. Additionally, we recommend implementing the diverse accessibility training recommendations outlined in Section \ref{sec:Recommendations} to assess their effectiveness in industry settings.

\subsection{Impact of this work}
The outcomes of our study carry significance for both the academic realm and the industry. Acquiring a substantial volume of survey responses from industry professionals is often a challenge, yet we successfully gathered insights from over 260 software professionals, establishing a robust foundation for our results.

The survey underwent meticulous design and refinement cycles, adhering to well-established scientific principles. Through the analysis of the survey data, we derived recommendations for enhancing accessibility education in the industry, offering valuable insights for future research and industry practitioners. Our work not only illuminates the state of digital accessibility in the Indian context but also raises awareness within the scientific community regarding this frequently overlooked topic. Additionally, we provide the raw survey data, facilitating further exploration of the subject.

% !TeX root = main.tex
\section{Conclusion}
\label{sec:conclusion}
Accessibility is increasingly critical in today's landscape, with the repercussions of neglecting it ranging from substantial financial penalties to the potential loss of business and harm to users with disabilities. In recent years, there has been a noticeable rise in accessibility-related legal actions. Despite the existence of accessibility standards and guidelines established by the W3C, their adherence by software professionals has fallen short, resulting in instances of inaccessibility. Given the prevailing circumstances, we delved into the following inquiries: (1) Who actively incorporates accessibility practices within the technology sector? (2) How and at what stages do professionals consider and implement accessibility? (3) What challenges do they face in ensuring accessibility? (4) What resources are essential for ensuring accessibility? Our extensive survey involving 269 software professionals in the Indian IT industry revealed that junior-level professionals exhibit lower engagement with accessibility topics than their senior counterparts. Additionally, 39\% of respondents practiced inclusive design but fell short of fully implementing its aspects. Challenges such as insufficient awareness and a scarcity of engineers skilled in accessibility were prominent. This emphasizes the urgent need for accessibility education in the industry, particularly in the Indian context. A substantial majority (69.9\%, N=269)
of respondents express the need for training materials, workshops, and bootcamps to enhance their accessibility skills. In response, we present a set of recommendations and best practices gathered from RQ4 to equip industry professionals with the requisite accessibility skills.

\section*{Data Availability}
The survey responses in the form of anonymized raw data have been uploaded as supplementary material in the submission process on the HotCRP submission site and will be freely accessible on Zenodo in CSV format (will be published on acceptance). Researchers are invited to utilize this dataset for additional studies.

\begin{acks}
  This work is supported by Birla Institute of Technology and Science, Pilani under grant GOA/ACG/2021-2022/Nov/05.
\end{acks}

\bibliographystyle{ACM-Reference-Format}
\bibliography{main.bib}


\begin{thebibliography}{25}

%%% the Web address.

\ifx \showCODEN    \undefined \def \showCODEN     #1{\unskip}     \fi
\ifx \showDOI      \undefined \def \showDOI       #1{#1}\fi
\ifx \showISBNx    \undefined \def \showISBNx     #1{\unskip}     \fi
\ifx \showISBNxiii \undefined \def \showISBNxiii  #1{\unskip}     \fi
\ifx \showISSN     \undefined \def \showISSN      #1{\unskip}     \fi
\ifx \showLCCN     \undefined \def \showLCCN      #1{\unskip}     \fi
\ifx \shownote     \undefined \def \shownote      #1{#1}          \fi
\ifx \showarticletitle \undefined \def \showarticletitle #1{#1}   \fi
\ifx \showURL      \undefined \def \showURL       {\relax}        \fi
% invisible to TeX
\providecommand\bibfield[2]{#2}
\providecommand\bibinfo[2]{#2}
\providecommand\natexlab[1]{#1}
\providecommand\showeprint[2][]{arXiv:#2}

\bibitem[w3c(2014)]%
        {w3cStartwitha11y}
 \bibinfo{year}{2014}\natexlab{}.
\newblock \bibinfo{title}{Start with Accessibility - Education \& Outreach}.
\newblock
\newblock
\urldef\tempurl%
\url{https://www.w3.org/WAI/EO/wiki/Start_with_Accessibility}
\showURL{%
\tempurl}


\bibitem[Sec(2018)]%
        {Section508}
 \bibinfo{year}{2018}\natexlab{}.
\newblock \bibinfo{booktitle}{\emph{Section 508 of the Rehabilitation Act of
  1973}}.
\newblock \bibinfo{type}{{T}echnical {R}eport} A/RES/61/106.
  \bibinfo{institution}{{United Nations}}.
\newblock


\bibitem[cva(2021)]%
        {cvaa_21st_2021}
 \bibinfo{year}{2021}\natexlab{}.
\newblock \bibinfo{title}{21st {Century} {Communications} and {Video}
  {Accessibility} {Act} ({CVAA})}.
\newblock
\newblock
\urldef\tempurl%
\url{https://www.fcc.gov/consumers/guides/21st-century-communications-and-video-accessibility-act-cvaa}
\showURL{%
\tempurl}


\bibitem[web(2021)]%
        {webAim_Pract_Survey}
 \bibinfo{year}{2021}\natexlab{}.
\newblock \bibinfo{title}{WebAIM: Survey of Web Accessibility Practitioners \#3
  Results}.
\newblock
\newblock
\urldef\tempurl%
\url{https://webaim.org/projects/practitionersurvey3/}
\showURL{%
\tempurl}


\bibitem[Lev(2022)]%
        {LevelAccess2022}
 \bibinfo{year}{2022}\natexlab{}.
\newblock \bibinfo{title}{2022 State of Digital Accessibility}.
\newblock
\newblock
\urldef\tempurl%
\url{https://www.levelaccess.com/wp-content/uploads/2022/10/2022_State_of_Digital_Accessibility_Report.pdf}
\showURL{%
\tempurl}


\bibitem[acc(2023)]%
        {accAct}
 \bibinfo{year}{2023}\natexlab{}.
\newblock \bibinfo{title}{Information and reservation services of carriers be
  accessible to individuals with visual, hearing, and other disabilities?}
\newblock
\newblock
\urldef\tempurl%
\url{https://www.ecfr.gov/current/title-14/part-382/section-382.43}
\showURL{%
\tempurl}


\bibitem[nas(2023)]%
        {nasscom2023}
 \bibinfo{year}{2023}\natexlab{}.
\newblock \bibinfo{title}{Technology Sector in India 2023 Strategic Review}.
\newblock
\newblock
\urldef\tempurl%
\url{https://nasscom.in/knowledge-center/publications/technology-sector-india-2023-strategic-review}
\showURL{%
\tempurl}


\bibitem[EU_(2023)]%
        {EU_web_2023}
 \bibinfo{year}{2023}\natexlab{}.
\newblock \bibinfo{title}{Web {Accessibility} {\textbar} {Shaping} {Europe}’s
  digital future}.
\newblock
\newblock
\urldef\tempurl%
\url{https://digital-strategy.ec.europa.eu/en/policies/web-accessibility}
\showURL{%
\tempurl}


\bibitem[Armas et~al\mbox{.}(2020)]%
        {armas_proposal_2020}
\bibfield{author}{\bibinfo{person}{Lisandra Armas}, \bibinfo{person}{Hesmeralda
  Rojas}, {and} \bibinfo{person}{Ronald Renteria}.}
  \bibinfo{year}{2020}\natexlab{}.
\newblock \showarticletitle{Proposal for an accessible software development
  model}. In \bibinfo{booktitle}{\emph{2020 3rd {International} {Conference} of
  {Inclusive} {Technology} and {Education} ({CONTIE})}}.
  \bibinfo{publisher}{IEEE}, \bibinfo{address}{Baja California Sur, Mexico},
  \bibinfo{pages}{104--109}.
\newblock
\showISBNx{978-1-72818-342-8}
\urldef\tempurl%
\url{https://doi.org/10.1109/CONTIE51334.2020.00028}
\showDOI{\tempurl}


\bibitem[Baker et~al\mbox{.}(2020)]%
        {baker_systematic_2020}
\bibfield{author}{\bibinfo{person}{Catherine~M. Baker},
  \bibinfo{person}{Yasmine~N. El-Glaly}, {and} \bibinfo{person}{Kristen
  Shinohara}.} \bibinfo{year}{2020}\natexlab{}.
\newblock \showarticletitle{A systematic analysis of accessibility in computing
  education research}. In \bibinfo{booktitle}{\emph{Proceedings of the 51st
  {ACM} technical symposium on computer science education}}
  \emph{(\bibinfo{series}{{SIGCSE} '20})}. \bibinfo{publisher}{Association for
  Computing Machinery}, \bibinfo{address}{New York, NY, USA},
  \bibinfo{pages}{107--113}.
\newblock
\showISBNx{978-1-4503-6793-6}
\urldef\tempurl%
\url{https://doi.org/10.1145/3328778.3366843}
\showDOI{\tempurl}


\bibitem[Bhatia et~al\mbox{.}(2023)]%
        {bhatia_integrating_2023}
\bibfield{author}{\bibinfo{person}{Jaskaran~Singh Bhatia},
  \bibinfo{person}{Parthasarathy P~D}, \bibinfo{person}{Snigdha Tiwari},
  \bibinfo{person}{Dhruv Nagpal}, {and} \bibinfo{person}{Swaroop Joshi}.}
  \bibinfo{year}{2023}\natexlab{}.
\newblock \showarticletitle{Integrating {Accessibility} in a {Mobile} {App}
  {Development} {Course}}. In \bibinfo{booktitle}{\emph{Proceedings of the 54th
  {ACM} {Technical} {Symposium} on {Computer} {Science} {Education} {V}. 1}}.
  \bibinfo{publisher}{ACM}, \bibinfo{address}{Toronto ON Canada},
  \bibinfo{pages}{1021--1027}.
\newblock
\showISBNx{978-1-4503-9431-4}
\urldef\tempurl%
\url{https://doi.org/10.1145/3545945.3569825}
\showDOI{\tempurl}


\bibitem[{Department of Higher Education, Government of India}(2021)]%
        {department_of_higher_education_government_of_india_all_2021}
\bibfield{author}{\bibinfo{person}{{Department of Higher Education, Government
  of India}}.} \bibinfo{year}{2021}\natexlab{}.
\newblock \bibinfo{title}{All {India} survey on {Higher} {Education} 2020-21}.
\newblock
\newblock
\urldef\tempurl%
\url{https://aishe.gov.in/aishe/viewDocument.action?documentId=322}
\showURL{%
\tempurl}


\bibitem[Durdu and Yerlikaya(2020)]%
        {durdu_perception_2020}
\bibfield{author}{\bibinfo{person}{Pınar~ONAY Durdu} {and}
  \bibinfo{person}{Zehra Yerlikaya}.} \bibinfo{year}{2020}\natexlab{}.
\newblock \showarticletitle{The {Perception} of {Website} {Accessibility}: {A}
  {Survey} of {Turkish} {Software} {Professionals}}.
\newblock \bibinfo{journal}{\emph{AJIT-e}} \bibinfo{volume}{11},
  \bibinfo{number}{41} (\bibinfo{date}{Aug.} \bibinfo{year}{2020}),
  \bibinfo{pages}{42--71}.
\newblock
\showISSN{1309-1581}
\urldef\tempurl%
\url{https://doi.org/10.5824/ajite.2020.02.003.x}
\showDOI{\tempurl}
\newblock
\shownote{Number: 41 Publisher: Akademik Bilişim Araştırmaları Derneği}.


\bibitem[{El-Glaly}(2020)]%
        {el-glalyTeachingAccessibilitySoftware2020}
\bibfield{author}{\bibinfo{person}{Yasmine~N. {El-Glaly}}.}
  \bibinfo{year}{2020}\natexlab{}.
\newblock \showarticletitle{Teaching Accessibility to Software Engineering
  Students}. In \bibinfo{booktitle}{\emph{Annual Conference on Innovation and
  Technology in Computer Science Education, {{ITiCSE}}}}.
\newblock
\showISSN{1942647X}
\urldef\tempurl%
\url{https://doi.org/10.1145/3328778.3366914}
\showDOI{\tempurl}


\bibitem[Jia et~al\mbox{.}(2021)]%
        {jiaInfusingAccessibilityProgramming2021}
\bibfield{author}{\bibinfo{person}{Lin Jia}, \bibinfo{person}{Yasmine~N.
  Elglaly}, \bibinfo{person}{Catherine~M. Baker}, {and}
  \bibinfo{person}{Kristen Shinohara}.} \bibinfo{year}{2021}\natexlab{}.
\newblock \showarticletitle{Infusing {{Accessibility}} into {{Programming
  Courses}}}. In \bibinfo{booktitle}{\emph{Extended {{Abstracts}} of the 2021
  {{CHI Conference}} on {{Human Factors}} in {{Computing Systems}}}}.
  \bibinfo{publisher}{{ACM}}, \bibinfo{address}{{Yokohama Japan}},
  \bibinfo{pages}{1--6}.
\newblock
\showISBNx{978-1-4503-8095-9}
\urldef\tempurl%
\url{https://doi.org/10.1145/3411763.3451625}
\showDOI{\tempurl}


\bibitem[Kawas et~al\mbox{.}(2019)]%
        {kawasTeachingAccessibilityDesign2019}
\bibfield{author}{\bibinfo{person}{Saba Kawas}, \bibinfo{person}{Laura
  Vonessen}, {and} \bibinfo{person}{Amy Ko}.} \bibinfo{year}{2019}\natexlab{}.
\newblock \showarticletitle{Teaching Accessibility: {{A}} Design Exploration of
  Faculty Professional Development at Scale}. In
  \bibinfo{booktitle}{\emph{{{SIGCSE}} 2019 - Proceedings of the 50th {{ACM}}
  Technical Symposium on Computer Science Education}}.
\newblock
\showISBNx{978-1-4503-5890-3}
\urldef\tempurl%
\url{https://doi.org/10.1145/3287324.3287399}
\showDOI{\tempurl}


\bibitem[Martin et~al\mbox{.}(2022)]%
        {martin_landscape_2022}
\bibfield{author}{\bibinfo{person}{Lilu Martin}, \bibinfo{person}{Catherine
  Baker}, \bibinfo{person}{Kristen Shinohara}, {and}
  \bibinfo{person}{Yasmine~N. Elglaly}.} \bibinfo{year}{2022}\natexlab{}.
\newblock \showarticletitle{The {Landscape} of {Accessibility} {Skill} {Set} in
  the {Software} {Industry} {Positions}}. In \bibinfo{booktitle}{\emph{The 24th
  {International} {ACM} {SIGACCESS} {Conference} on {Computers} and
  {Accessibility}}}. \bibinfo{publisher}{ACM}, \bibinfo{address}{Athens
  Greece}, \bibinfo{pages}{1--4}.
\newblock
\showISBNx{978-1-4503-9258-7}
\urldef\tempurl%
\url{https://doi.org/10.1145/3517428.3550389}
\showDOI{\tempurl}


\bibitem[{Ministry of Social Justice and Empowerment, Government of
  India}(2016)]%
  {ministryofsocialjusticeandempowermentgovernmentofindiaRightsPersonsDisabilities2016}
\bibfield{author}{\bibinfo{person}{{Ministry of Social Justice and Empowerment,
  Government of India}}.} \bibinfo{year}{2016}\natexlab{}.
\newblock \bibinfo{title}{The {{Rights}} of {{Persons}} with {{Disabilities}}
  ({{RPD}}) {{Act}}, 2016}.
\newblock
\newblock


\bibitem[Patel et~al\mbox{.}(2020)]%
        {patel_why_2020}
\bibfield{author}{\bibinfo{person}{Rohan Patel}, \bibinfo{person}{Pedro
  Breton}, \bibinfo{person}{Catherine~M. Baker}, \bibinfo{person}{Yasmine~N.
  El-Glaly}, {and} \bibinfo{person}{Kristen Shinohara}.}
  \bibinfo{year}{2020}\natexlab{}.
\newblock \showarticletitle{Why {Software} is {Not} {Accessible}: {Technology}
  {Professionals}' {Perspectives} and {Challenges}}. In
  \bibinfo{booktitle}{\emph{Extended {Abstracts} of the 2020 {CHI} {Conference}
  on {Human} {Factors} in {Computing} {Systems}}}. \bibinfo{publisher}{ACM},
  \bibinfo{address}{Honolulu HI USA}, \bibinfo{pages}{1--9}.
\newblock
\showISBNx{978-1-4503-6819-3}
\urldef\tempurl%
\url{https://doi.org/10.1145/3334480.3383103}
\showDOI{\tempurl}


\bibitem[Pool(2023)]%
        {pool_accessibility_2023}
\bibfield{author}{\bibinfo{person}{Jonathan~Robert Pool}.}
  \bibinfo{year}{2023}\natexlab{}.
\newblock \showarticletitle{Accessibility {Metatesting}: {Comparing} {Nine}
  {Testing} {Tools}}. In \bibinfo{booktitle}{\emph{20th {International} {Web}
  for {All} {Conference}}}. \bibinfo{publisher}{ACM}, \bibinfo{address}{Austin
  TX USA}, \bibinfo{pages}{1--4}.
\newblock
\showISBNx{9798400707483}
\urldef\tempurl%
\url{https://doi.org/10.1145/3587281.3587282}
\showDOI{\tempurl}


\bibitem[Richards et~al\mbox{.}(2012)]%
        {richards_web_2012}
\bibfield{author}{\bibinfo{person}{John~T. Richards}, \bibinfo{person}{Kyle
  Montague}, {and} \bibinfo{person}{Vicki~L. Hanson}.}
  \bibinfo{year}{2012}\natexlab{}.
\newblock \showarticletitle{Web accessibility as a side effect}. In
  \bibinfo{booktitle}{\emph{Proceedings of the 14th international {ACM}
  {SIGACCESS} conference on {Computers} and accessibility}}.
  \bibinfo{publisher}{ACM}, \bibinfo{address}{Boulder Colorado USA},
  \bibinfo{pages}{79--86}.
\newblock
\showISBNx{978-1-4503-1321-6}
\urldef\tempurl%
\url{https://doi.org/10.1145/2384916.2384931}
\showDOI{\tempurl}


\bibitem[Saldaña(2021)]%
        {saldana_coding_2021}
\bibfield{author}{\bibinfo{person}{Johnny Saldaña}.}
  \bibinfo{year}{2021}\natexlab{}.
\newblock \bibinfo{booktitle}{\emph{The coding manual for qualitative
  researchers} (\bibinfo{edition}{4e [fourth editiion]} ed.)}.
\newblock \bibinfo{publisher}{SAGE Publishing Ltd}, \bibinfo{address}{London
  [England] ; Thousand Oaks}.
\newblock
\showISBNx{978-1-5297-3174-3 978-1-5297-3175-0}


\bibitem[Secretariat(2011)]%
        {secretariat_standard_2011}
\bibfield{author}{\bibinfo{person}{Treasury Board of~Canada Secretariat}.}
  \bibinfo{year}{2011}\natexlab{}.
\newblock \bibinfo{title}{Standard on {Web} {Accessibility}}.
\newblock
\newblock
\urldef\tempurl%
\url{https://www.tbs-sct.canada.ca/pol/doc-eng.aspx?id=23601}
\showURL{%
\tempurl}
\newblock
\shownote{Last Modified: 2011-08-01}.


\bibitem[Tseng et~al\mbox{.}(2022)]%
        {tseng_exploration_2022}
\bibfield{author}{\bibinfo{person}{Chia-En Tseng}, \bibinfo{person}{Seoung~Ho
  Jung}, \bibinfo{person}{Yasmine~N. Elglaly}, \bibinfo{person}{Yudong Liu},
  {and} \bibinfo{person}{Stephanie Ludi}.} \bibinfo{year}{2022}\natexlab{}.
\newblock \showarticletitle{Exploration on {Integrating} {Accessibility} into
  an {AI} {Course}}. In \bibinfo{booktitle}{\emph{Proceedings of the 53rd {ACM}
  {Technical} {Symposium} on {Computer} {Science} {Education}}}.
  \bibinfo{publisher}{ACM}, \bibinfo{address}{Providence RI USA},
  \bibinfo{pages}{864--870}.
\newblock
\showISBNx{978-1-4503-9070-5}
\urldef\tempurl%
\url{https://doi.org/10.1145/3478431.3499399}
\showDOI{\tempurl}


\bibitem[Yan and Ramachandran(2019)]%
        {yan_current_2019}
\bibfield{author}{\bibinfo{person}{Shunguo Yan} {and} \bibinfo{person}{P.~G.
  Ramachandran}.} \bibinfo{year}{2019}\natexlab{}.
\newblock \showarticletitle{The {Current} {Status} of {Accessibility} in
  {Mobile} {Apps}}.
\newblock \bibinfo{journal}{\emph{ACM Trans. Access. Comput.}}
  \bibinfo{volume}{12}, \bibinfo{number}{1} (\bibinfo{date}{Feb.}
  \bibinfo{year}{2019}), \bibinfo{pages}{1--31}.
\newblock
\showISSN{1936-7228, 1936-7236}
\urldef\tempurl%
\url{https://doi.org/10.1145/3300176}
\showDOI{\tempurl}


\end{thebibliography}

% Uncomment if required.

\end{document}